\documentclass[10pt,conference]{IEEEtran}
\usepackage{amsmath,amsfonts}
\usepackage{algorithmic}
\usepackage{array}
\usepackage[caption=false,font=normalsize,labelfont=sf,textfont=sf]{subfig}
\usepackage[T1]{fontenc}
\newcolumntype{P}[1]{>{\centering\arraybackslash}p{#1}}
\usepackage{textcomp}
\usepackage{stfloats}
\usepackage{url}
\usepackage{verbatim}
\usepackage{xcolor}
\usepackage{graphicx}
\usepackage{booktabs}
\usepackage{tabularx}
\usepackage{makecell}
\usepackage{multirow}
\usepackage{multicol}
\usepackage{hyperref}
\usepackage{moresize}
\def\BibTeX{{\rm B\kern-.05em{\sc i\kern-.025em b}\kern-.08em
    T\kern-.1667em\lower.7ex\hbox{E}\kern-.125emX}}
\usepackage{balance}

\begin{document}

\title{Vulnerability Assessment Combining CVSS Temporal Metrics and Bayesian Networks}

\author{\IEEEauthorblockN{Stefano Perone, Simone Guarino, Luca Faramondi, Roberto Setola}\\
\IEEEauthorblockA{Unit of Automatic Control,\\
Department of Engineering,\\
Campus Bio-Medico University of Rome,\\
Via \'Alvaro del Portillo 21, 00128 Rome, Italy\\
Email: \{s.perone, s.guarino, l.faramondi, r.setola\}@unicampus.it}
}

\maketitle

\begin{abstract}
Vulnerability assessment is a critical challenge in cybersecurity, particularly in industrial environments. This work presents an innovative approach by incorporating the temporal dimension into vulnerability assessment, an aspect neglected in existing literature. Specifically, this paper focuses on refining vulnerability assessment and prioritization by integrating Common Vulnerability Scoring System (CVSS) Temporal Metrics with Bayesian Networks to account for exploit availability, remediation efforts, and confidence in reported vulnerabilities. Through probabilistic modeling, Bayesian networks enable a structured and adaptive evaluation of vulnerabilities, allowing for more accurate prioritization and decision-making. The proposed approach dynamically computes the Temporal Score and updates the CVSS Base Score by processing data on exploits and fixes from vulnerability databases. 
In the case study, we apply this approach to an industrial infrastructure modeled using the Purdue Model. The dependencies among vulnerabilities are then modeled through a Bayesian Attack Graph (BAG) to compute the posterior exploitation probabilities at each BAG node.
\end{abstract}


\section{Introduction}
In the contemporary cybersecurity environment, the need for a mechanism that can identify, categorize, and prioritize threats and vulnerabilities in systems is increasingly apparent. Companies and organizations must comprehend the risks associated with specific systems and formulate remediation and mitigation strategies. The most prevalent tool employed for assessing vulnerabilities is the Common Vulnerability Scoring System (CVSS) \cite{CVSS_original}, a \textit{de facto} standard associated with the respective Common Vulnerabilities and Exposures (CVE). The CVE is designed to uniquely identify each vulnerability.

CVSS captures the main technical characteristics of software, hardware, and firmware vulnerabilities \cite{CVSS_docs}, and its outputs include numerical scores that indicate the severity of the considered vulnerability.
CVSS version 3.1 is composed of three metric groups: Base, Temporal, and Environmental. The Base metric group represents the intrinsic characteristics of a vulnerability that are constant over time and across user environments. The Temporal Metrics include measurements of the present state of exploit techniques or code availability, the existence of any patches or workarounds, and the confidence in the description of a vulnerability. Environmental metrics empower analysts to personalize the CVSS score, considering the significance of the compromised asset within an organization's context. This is measured in terms of complementary/alternative security controls in place, as well as Confidentiality, Integrity, and Availability.

The purpose of the paper is to make the vulnerability analysis as specific and contextualized as possible. A novel approach to achieving this objective entails the assessment of the temporal dimension. A vulnerability that is deemed critical in the present may potentially diminish in severity in the future, possibly attributable to the development of a solution or a reduction in its susceptibility to exploitation. To this end, the paper proposes the utilization of CVSS Temporal Metrics for the evaluation of the impact of time on vulnerabilities, to prioritize the vulnerabilities of a system and analyze how these metrics influence vulnerability exploitation probability. These objectives will be accomplished through a preliminary analysis of vulnerability databases, collecting data about exploits and fixes associated with a specific vulnerability, to compute CVSS Temporal Metrics and update the CVSS Base Score. Then, we employ Bayesian Networks (BN), a class of probabilistic graphical models represented by directed acyclic graphs $\mathcal{G} = \{\mathcal{V}, \mathcal{E}\}$, where $v_i \in \mathcal{V}$ represents a random variable $X_i$ and each edge $(v_i, v_j) \in \mathcal{E}$ represents causal relationships among nodes \cite{guarino1}. The value of a BN node that is known \textit{a priori} is referred to as evidence. Upon receiving new evidence, a BN computes the probabilities for all the other nodes using the Conditional Probability Tables (CPTs) \cite{guarino3}.
In detail, a CPT specifies the conditional probability distributions of the descendant, the child node, for all combinations of the states of its direct predecessors, the parent nodes \cite{laitila}.
To enhance the visualization of the network and its potential attack paths, we employ a Bayesian Attack Graph (BAG), a tailored adaptation of BNs designed to capture both attacker behavior and the progression of attacks throughout the network. Specifically, nodes represent compromising events, while edges represent conditional relationships between nodes. The information available at each node in the BAG is the conditional probability distribution $p(X_i|{p_a}_i)$, i.e. the probability of a node $X_i$ to be compromised given the state of its parent nodes ${p_a}_i$ \cite{Gonzalez}.

The proposed approach is applied to a case study consisting of an industrial network infrastructure modeled through the Purdue Enterprise Reference Architecture \cite{williams}. We first identify system vulnerabilities and then apply the vulnerability analysis approach by computing CVSS Temporal Scores and evaluating the probabilities of compromising events through a BAG. 

The rest of the paper is organized as follows. Section \ref{sec:related_works} provides an overview of the related literature. Section \ref{sec:approach} presents the proposed vulnerability analysis approach. Section \ref{sec:case_study} describes an application of the proposed approach to the case study, while Section \ref{sec:conclusion} concludes the paper.

\section{Related works} \label{sec:related_works}
For what concerns the utilization of BNs in the context of vulnerability assessment, this topic is undergoing significant expansion and development. This is well-explained by George et al. \cite{george}, who examined the evolution of security risk assessment methodologies in industrial sectors presenting BNs as an emerging tool for dynamically assessing risks by integrating new information and handling multi-state variables. In this context, Huang et al. \cite{huang} proposed a BN-based risk assessment model for SCADA systems that aims to dynamically and quantitatively assess the security risk level and integrates traditional BN-based security risk assessment methods with the Leaky Noisy-OR gate to support risk assessment for unknown attacks. 
Guarino et al. \cite{guarino2} proposed a new risk assessment approach that combines BNs with multi-criteria decision-making to provide a holistic risk metric by integrating several heterogeneous risk values calculated through BNs.
Similarly, Poolsappasit et al. \cite{Poolsappasit} proposed a dynamic risk management framework utilizing Bayesian Attack Graphs, which captures the causal dependencies between various network states and vulnerabilities. This approach addressed the limitations of traditional risk assessment models by incorporating Bayesian logic to quantify the likelihood of different attack paths. By doing so, the framework enables system administrators to not only analyze potential threats but also to dynamically adjust their security strategies in response to evolving conditions within the network.
Singh et al. \cite{singh} proposed a novel risk estimation model that utilizes CVSS and CVE data to evaluate security risks. Their approach combines the frequency of vulnerabilities and the maturity of exploit code to generate a comprehensive assessment of risk levels associated with vulnerabilities in a network. By focusing on the Temporal and Environmental vectors of the CVSS score, the proposed model allows for an understanding of vulnerability impact.
Meng et al. \cite{meng}, instead, directed their attention toward the prioritization of physical risks, proposing a hybrid Bayesian Network model that integrates physical knowledge with data-driven learning to prioritize risk-influencing factors (RIFs). This model combines structure learning and parameter learning to identify causal relationships among RIFs, providing valuable insights for emergency planning and risk mitigation strategies.
Sato et al. \cite{sato} proposed the Exploit Time Probability (ETP)-model, which emphasizes the significant impact of time on the exploit probabilities of vulnerabilities.
They focused on the elapsed time since the vulnerability was published rather than on CVSS metrics.
Their analysis revealed that the exploit time distribution is concentrated around zero, indicating that vulnerabilities are often exploited shortly before or after their publication in the National Vulnerability Database (NVD), but giving poor information about the effective impact of time as described in CVSS Temporal Metrics.

However, these works fail to assess the impact of time on vulnerabilities, except for \cite{singh} and \cite{sato} which solely consider the elapsed time since the publication of a CVE to calculate exploitation probability. This approach is limited in scope as it does not take into account a broader temporal analysis across different data sources. In contrast, our approach is innovative as it incorporates a comprehensive evaluation of the impact of time by analyzing multiple vulnerability databases. This allows us to capture a more holistic view of how vulnerabilities evolve, their severity over time, and how their risk factors change in different contexts. By integrating several data sources, we provide a deeper understanding of vulnerability trends, making our approach more robust and insightful compared to previous studies.

\section{Vulnerability Analysis Approach} \label{sec:approach}
The primary objective of this study is to assess and prioritize vulnerabilities within a system by leveraging CVSS Temporal Metrics and Bayesian Networks. 
\begin{figure}[ht!]
\includegraphics[scale=.68]{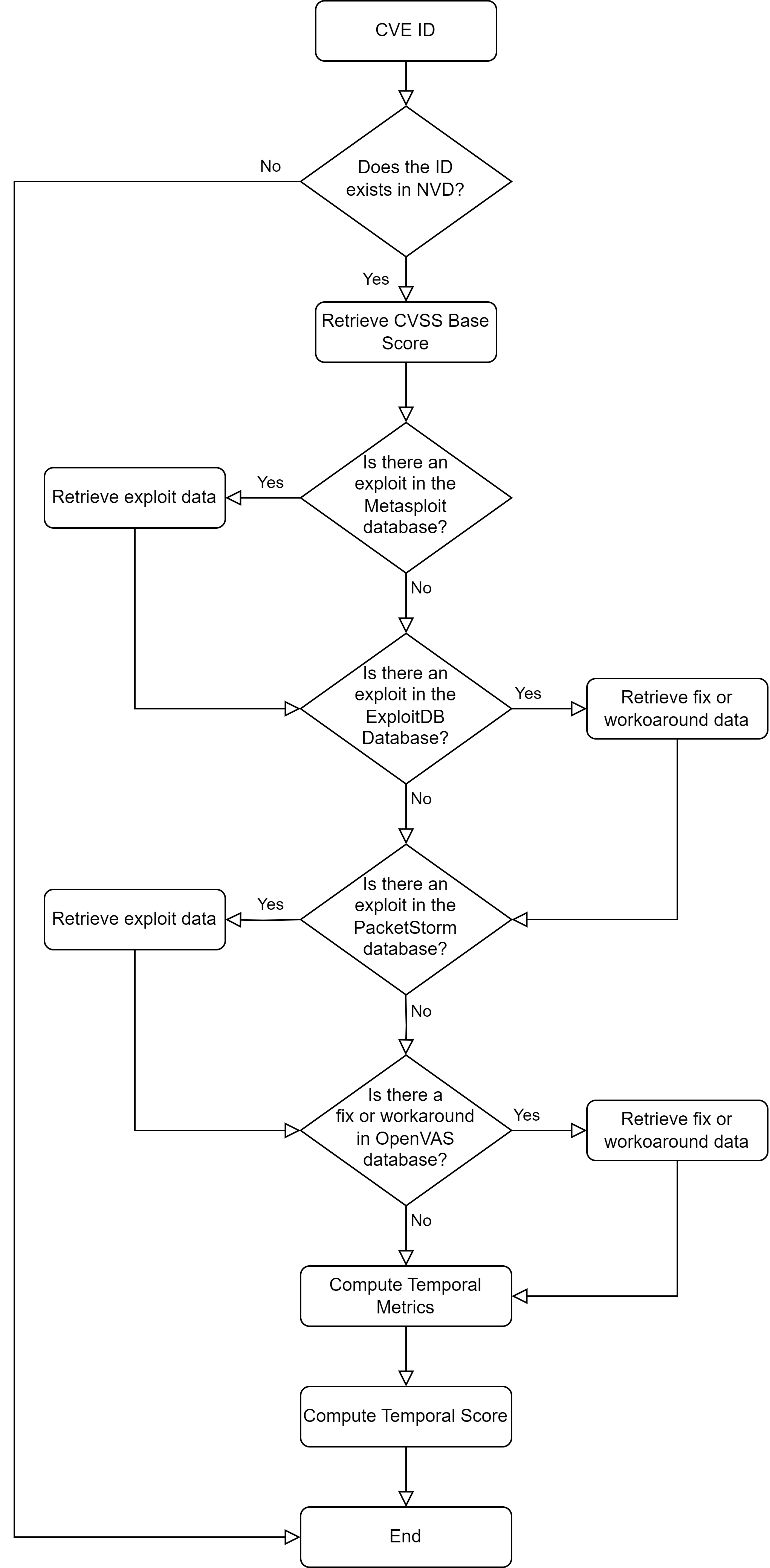}
\caption{Flowchart of the realized script.}
\label{fig:flowchart}
\end{figure}
According to the CVSS version 3.1 standard, three primary Temporal Metrics influence the vulnerability score over time:

\begin{itemize}
    \item \textbf{Exploit Code Maturity (ECM):} measures the likelihood of the vulnerability being attacked and is typically based on the current state of exploit techniques, exploit code availability, or active, “in-the-wild” exploitation \cite{CVSS_docs}.
    \item \textbf{Remediation Level (RL):} indicates the availability of either official patches or workarounds, as provided by vendors, or alternative solutions. It highlights the level of mitigation for a vulnerability.
    \item \textbf{Report Confidence (RC):} measures the degree of confidence in the existence of the vulnerability and the credibility of the known technical details.
\end{itemize}
The possible values and corresponding numerical scores for these metrics are summarized in Table \ref{table:values}.

To compute the CVSS Temporal Metrics and the updated vulnerability score, a dedicated script is developed that retrieves CVE data and verifies the presence of exploits and fixes. Figure \ref{fig:flowchart} presents a detailed flowchart outlining the script's operation.

\begin{table}[!htb]
\renewcommand{\thetable}{\arabic{table}}
\renewcommand{\arraystretch}{1.5}
\centering
\caption{Values of Temporal Metrics.}
\label{table:values}
\resizebox{.485\textwidth}{!}{
\begin{tabular}{c c c}
\toprule
\textbf{Metric}&\textbf{Metric Value}&\textbf{Numerical Value}\\
\midrule
& Not Defined (X) & 1 \\
& High (H) & 1 \\
Exploit Code Maturity & Functional (F) & 0.97 \\
& Proof-of-Concept (P) & 0.94 \\
& Unproved (U) & 0.91 \\
\midrule
& Not Defined (X) & 1 \\
& Unavailable (U) & 1 \\
Remediation Level & Workaround (W) & 0.97 \\
& Temporary Fix (T) & 0.96 \\
& Official Fix (O) & 0.95 \\
\midrule
& Not Defined (X) & 1 \\
Report & Confirmed (C) & 1 \\
Confidence & Reasonable (R) & 0.96 \\
& Unknown (U) & 0.92 \\
\bottomrule
\end{tabular}}
\end{table}

The script takes as input a list of CVEs and begins to query the NVD\footnote{\href{https://nvd.nist.gov/}{ https://nvd.nist.gov}}, which is maintained by the National Institute of Standards and Technology (NIST). The NVD is the primary repository for CVEs and provides detailed information on vulnerabilities, including their CVSS Base Score. The script sends a request to the NVD API using the CVE ID provided as input. If the CVE exists, the script retrieves the CVSS Base Score related to that vulnerability.

Subsequently, the script searches for the existence of exploits related to the vulnerability under examination. The script assesses the likelihood of exploitation by searching for known exploits across three major exploit databases.
The initial database examined is the Metasploit\footnote{\href{https://www.metasploit.com/}{https://www.metasploit.com}} database. Metasploit is a widely used penetration testing framework that provides detailed information about exploits, including their effectiveness. The script queries the Metasploit database and retrieves: \textit{name}, \textit{description}, and \textit{rank}. The latter field is of particular importance as it allows for the categorization of the severity of the exploit and the evaluation of the Exploit Code Maturity. In particular, the script assigns the ECM value as described below:

\begin{itemize}
    \item High (H): if the rank has a value as \textit{Excellent} or \textit{Great}
    \item Functional (F): if the rank has a value as \textit{Good} or \textit{Normal}
    \item Proof-of-Concept (P): otherwise
\end{itemize}

The second database examined is the Exploit Database\footnote{\href{https://www.exploit-db.com/}{https://www.exploit-db.com}}; it is a public repository of exploits for various vulnerabilities. In this case, the following information is retrieved: \textit{id}, \textit{description}, \textit{date\_published}, \textit{author}, \textit{type}, \textit{platform}, \textit{url} and \textit{verified}. As with the Metasploit Database, the latter field is used to assess the ECM value. Specifically, the field \textit{verified} serves to indicate whether an exploit has been tested and confirmed to function as intended. When set to 1 (true), the ECM is designated as "Functional (F)". Conversely, when set to 0 (false), the assigned value is "Proof-of-Concept (P)".

The final exploit database examined is that of PacketStorm\footnote{\href{https://www.packetstormsecurity.com/}{https://www.packetstormsecurity.com}}. PacketStorm provides a collection of exploits but with limited metadata; so, in this instance, it is only feasible to ascertain the presence or absence of an exploit pertaining to the vulnerability. In the event of a positive identification, the ECM is set to "Proof-of-Concept (P)".

The final ECM value assigned to the vulnerability is the maximum among the values of the three databases.

In the event that exploits corresponding to the vulnerability under consideration are not identified in all three of the databases, the ECM value is set to "Unproven (U)".

After verifying the presence of exploits, the script proceeds to research for the presence of fixes or workarounds related to the vulnerability under exam in the OpenVAS\footnote{\href{https://www.openvas.org/}{ https://www.openvas.org/}} database, which contains information on security solutions for known vulnerabilities. In this case, the information retrieved is only represented by the field \textit{solution\_type}, which refers to the type of remediation or mitigation available for the vulnerability.
In detail, the script makes conversions from OpenVAS values to RL values as described below:

\begin{itemize}
    \item \textit{NoneAvailable} $\rightarrow$ \textit{Unavailable (U)}
    \item \textit{WillNotFix} $\rightarrow$ \textit{Unavailable (U)}
    \item \textit{Workaround} $\rightarrow$ \textit{Temporary Fix (T)}
    \item \textit{Mitigation} $\rightarrow$ \textit{Workaround (W)}
    \item \textit{VendorFix} $\rightarrow$ \textit{Official Fix (O)}
\end{itemize}

It should be noted that the assignment of the "Workaround" value of the OpenVAS database to the "Temporary Fix (T)" value of the Remediation Level is due to the definition of "Workaround" in OpenVAS, which indicates a temporary solution to avoid a vulnerability, while in \cite{CVSS_docs} it is defined as an unofficial, non-vendor solution available.

Finally, the last action of the script is to compute the Temporal Score, that is the CVSS Base Score weighted with the Temporal Metrics.
In detail, the Temporal Score is computed using the following formula:
{
\small
\[TemporalScore = RoundUp(BaseScore \, * \, ECM \, * \, RL \, * \, RC)\]
}
where \textit{RoundUp} is a function that approximates the result to the first decimal digit.

\section{Case study} \label{sec:case_study}
In this section, we describe a case study in which we apply the proposed approach to analyze vulnerabilities within a simulated Industrial Control System (ICS). The goal is to demonstrate how integrating CVSS Temporal Metrics with Bayesian Networks can enhance vulnerability assessment and prioritization.
\subsection{Setup}
\begin{figure}[ht!]
\includegraphics[scale=.34]{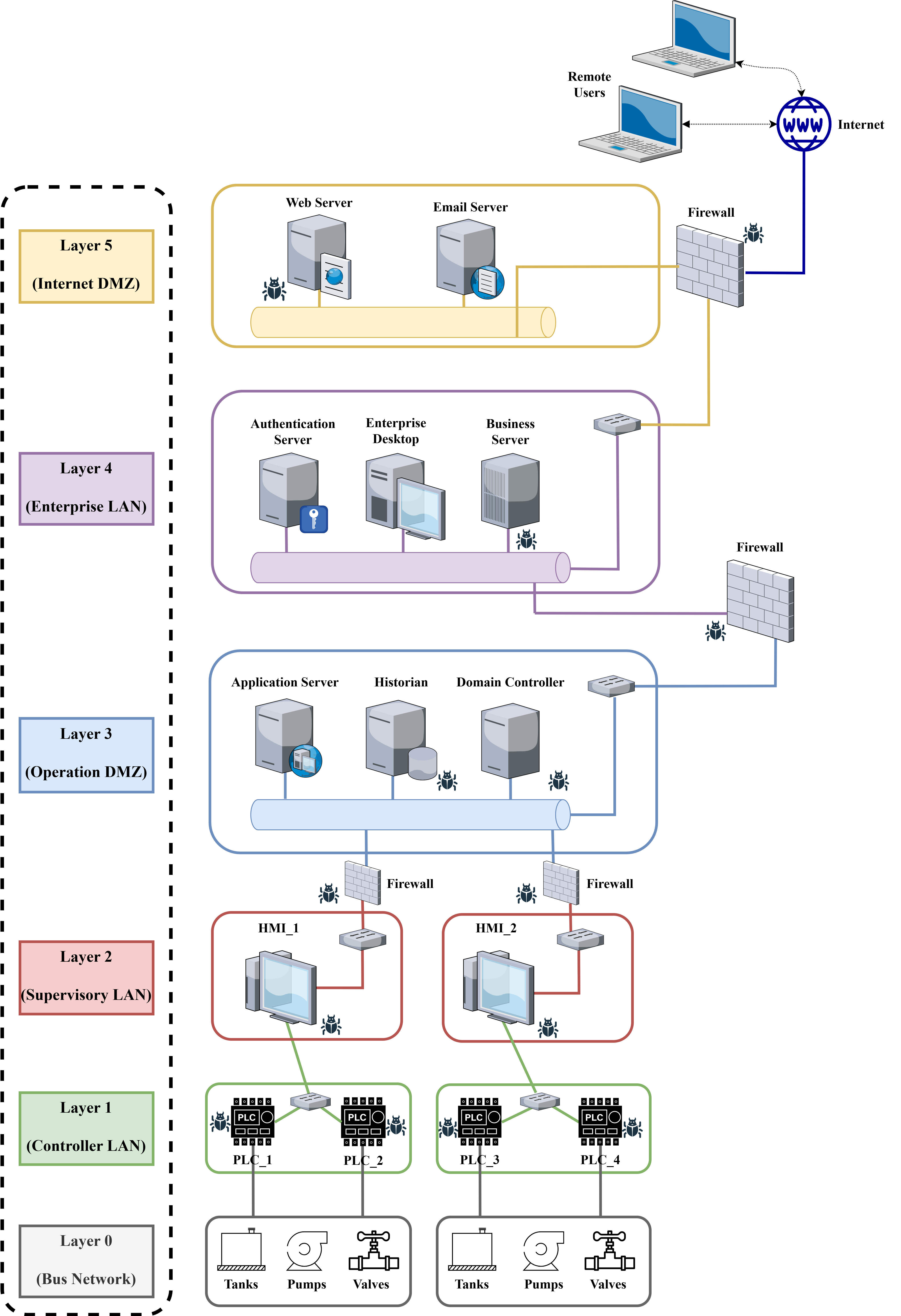}
\caption{ICS network infrastructure organized based on the Purdue Model.}
\label{fig:purdue}
\end{figure}

Initially, we structure the ICS network infrastructure into multiple layers following the Purdue Model, as shown in Figure \ref{fig:purdue}.

The Purdue Model defines several key layers of an attack:

\begin{itemize}
    \item \textbf{Layer 5 (Perimeter Security)} – The attacker breaches the network through a firewall vulnerability, bypassing traditional security defenses.
    \item \textbf{Layer 4 (Web and Business Servers)} – By compromising a web server, the attacker gains root access and further infiltrates a business server.
    \item \textbf{Layer 3 (Domain and Network Infrastructure)} - The attack spreads deeper by targeting domain controllers and firewalls, allowing lateral movement within the internal network.
    \item \textbf{Layers 1-2 (Industrial Control Systems)} - The attacker breaches Human-Machine Interfaces (HMIs), enabling direct access to Programmable Logic Controllers (PLCs).
\end{itemize}

\begin{figure}[ht!]
\includegraphics[scale=.4]{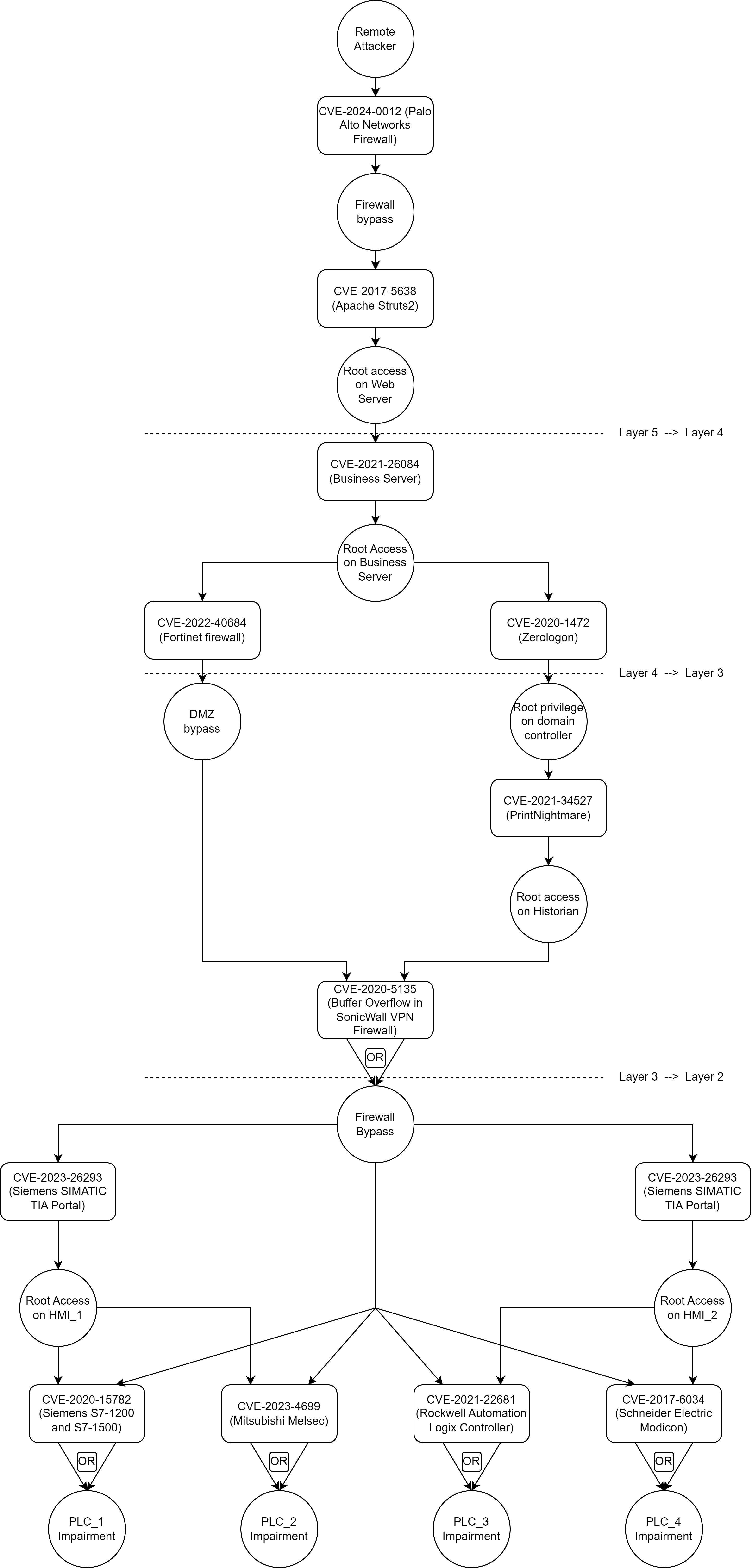}
\caption{Bayesian Attack Graph of the simulated system.}
\label{fig:bag}
\end{figure}

In the analysis, we identify specific vulnerabilities in selected systems within each layer of the Purdue Model. These discovered vulnerabilities are used to construct the BAG shown in Figure \ref{fig:bag}.
The BAG allows to visualize the possible attack paths that an attacker could take to reach the industrial control devices at Layer 1, based on the exploitation of these vulnerabilities.

This layered model, along with the analysis of identified vulnerabilities and the BAG, allows us to map potential attack vectors and assess their probabilities. It highlights how the exploitation of vulnerabilities at higher layers (e.g., firewalls and web servers) can cascade into deeper, more critical areas (such as PLCs). Actually, the BAG illustrates the potential real-world consequences of cyber intrusions, ranging from data breaches to industrial sabotage.



Each BAG node’s CPT is computed based on the score associated with the corresponding vulnerability. In more detail, the CPTs describe the probability that one node may be compromised given the security state of the parent nodes and the specific vulnerability affecting it. Table \ref{table:cpt} shows the structure of the CPT associated with node i-$th$, where the probability ${P(X_i)}$ is computed based on the following formula: $P(X_i) = Score / 10$, as the vulnerability score is defined between 0 and 10. In our evaluation, we compare the Bayesian inference results by first assigning the Score value using the Base Score and then using the computed Temporal Score, as detailed in Section \ref{sec:approach}.



\begin{table}[h!]
\renewcommand{\thetable}{\arabic{table}}
\renewcommand{\arraystretch}{1.5}
\centering
\caption{Conditional Probability Table Schema.}
\label{table:cpt}
\begin{tabular}{c|c|c|c}
\cmidrule{3-4}
\multicolumn{2}{c}{} & \multicolumn{2}{c}{\textbf{Inference}} \\ \cmidrule{3-4}
\multicolumn{2}{c}{} & True & False \\ \midrule
\multirow{2}{*}{\textbf{Evidence}} & True & $P(X_i)$ & $1 - P(X_i)$ \\ 
 & False & $0$ & $1$\\  
 \bottomrule
\end{tabular}
\end{table}


\subsection{Analysis and Evaluation}



The objective of this subsection is to illustrate the calculated Temporal Score value for each vulnerability and to compute the probability of node exploitation based on the score value. This shows the difference depending on whether the Base Score or Temporal Score is considered and shows the influence of Temporal Metrics on the exploitation probability value.

Table \ref{table:temporal} shows the Temporal Score computed for each vulnerability and the comparison with the corresponding Base Score.

\begin{table}[!htb]
\fontsize{7pt}{7pt}\selectfont
\renewcommand{\thetable}{\arabic{table}}
\renewcommand{\arraystretch}{1.5}
\centering
\caption{Temporal Scores of system vulnerabilities.}
\label{table:temporal}
\begin{tabular}{c c c}
\toprule
\textbf{CVE ID}&\textbf{CVSS Base Score}&\textbf{Temporal Score}\\
\midrule
CVE-2024-0012 & 9.8 & 8.9 \\
CVE-2017-5638 & 9.8 & 9.3 \\
CVE-2021-26084 & 9.8 & 9.8 \\
CVE-2022-40684 & 9.8 & 9.8 \\
CVE-2020-1472 & 10.0 & 8.9 \\
CVE-2021-34527 & 8.8 & 8.2 \\
CVE-2020-5135 & 9.8 & 8.9 \\
CVE-2023-26293 & 7.8 & 7.1 \\
CVE-2020-15782 & 9.8 & 8.9 \\
CVE-2023-4699 & 10.0 & 9.1 \\
CVE-2021-22681 & 9.8 & 8.9 \\
CVE-2017-6034 & 9.8 & 8.9 \\
\bottomrule
\end{tabular}
\end{table}

A key observation from these results is that all Temporal Score values are lower than or equal to their respective Base Scores. This phenomenon occurs because Temporal Metrics scale the score value downward, given their numerical values ranging from 0.91 to 1. This mechanism enables a more realistic and dynamic assessment, thereby avoiding the overestimation of vulnerabilities that, while critical at the time of discovery, may have diminished in dangerousness over time.

Subsequently, the exploitation probabilities at each node are computed based on both the Base and Temporal Scores. This analysis resulted in findings indicating a substantial impact of Temporal Scores on the calculation of inference, with discrepancies in some cases reaching over 20\%.

Table \ref{table:inferences} shows the exploitation probability at PLC nodes considering five different evidence scenarios.
\begin{table}[!htb]
\fontsize{7pt}{7pt}\selectfont
\setlength{\tabcolsep}{1pt}
\renewcommand{\thetable}{\arabic{table}}
\renewcommand{\arraystretch}{1.5}
\centering
\caption{Impact of Temporal Metrics on \\ vulnerability exploitation probability.}
\label{table:inferences}
\begin{tabular}{P{3cm} P{2.5cm} P{3cm}}
\toprule
\textbf{Inference} & \makecell{\textbf{Probability} \\ \textbf{(BN with Base Scores)}} & \makecell{\textbf{Probability} \\ \textbf{(BN with Temporal Scores)}}\\
\toprule
{\ttfamily P(PLC\_2 Impairment)} & 36.5\% & 25.1\% \\
\midrule
{\ttfamily P(PLC\_2 Impairment | Remote attacker)} & 73.0\% & 50.1\%\\
\midrule
{\ttfamily P(PLC\_3 Impairment | DMZ Bypass)} & 76.3\% & 60.9\%\\
\midrule
{\ttfamily P(PLC\_1 Impairment | Root access on Historian)} & 76.4\% & 62.4\% \\
\midrule
{\ttfamily P(PLC\_4 Impairment | Root access on HMI\_2)} & 98.0\% & 89.0\% \\
\bottomrule
\end{tabular}
\end{table}
The results indicate a significant difference in the probabilities obtained when using the Temporal Score compared to the Base Score. Specifically:
\begin{itemize}
    \item In the first scenario, the marginal probability of compromising PLC\_2 without evidence about the presence of an attacker is evaluated. The probability value drops from 36.5\% (Base Score) to 25.1\% (Temporal Score). This reduction reflects the fact that the associated vulnerabilities have lower values due to the downscale of all CVSS Base Scores with the influence of Temporal Metrics.
    \item In the second case, we analyzed the probability of compromising PLC\_2 with evidence about the presence of an attacker. This scenario reveals the most significant decrease, from 73.0\% to 50.1\%. 
    \item In the third scenario, the probability of compromising PLC\_3 decreases from 76.3\% to 60.9\% because several vulnerabilities in the previous nodes have vendor patches, lowering the RL values and thus reducing the overall risk.
    \item In the fourth example, the probability of compromising PLC\_1 drops from 76.4\% to 62.4\%. Despite a \textit{Functional (F)} level of ECM, the reduction is due to the availability of a temporary fix that mitigates the risk.
    \item In the end, the probability of compromising PLC\_4 decreases from 98.0\% to 89.0\% due to the unproven existence of exploits for all previous nodes of the attack.
\end{itemize}

These results highlight how Temporal Metrics offer a more realistic and time-sensitive assessment of vulnerability exploitation probability, enabling organizations to prioritize their responses based on current threat conditions.


\section{Conclusion} \label{sec:conclusion}
In this work, we presented a novel approach for vulnerability assessment and prioritization that combines the Common Vulnerability Scoring System (CVSS) Temporal Metrics with Bayesian Attack Graphs (BAG), demonstrating how a probabilistic framework could enhance cybersecurity strategies. By leveraging up-to-date exploitation data, remediation efforts, and expert confidence levels, our approach moves beyond static, baseline assessments and offers a dynamic mechanism to account for the evolving nature of threats.

Our case study demonstrates that when vulnerabilities are re-scored using CVSS Temporal Metrics instead of static CVSS Base Scores, organizations can achieve a more realistic and dynamic ranking. As shown in our results, CVSS Base Scores alone might overestimate threat levels of vulnerabilities that have been patched, are hard to exploit, or carry questionable credibility. Conversely, vulnerabilities with confirmed exploits and slow or non-existent remediation can rise in priority, alerting security teams to potential blind spots in their current patch-management processes.
Moreover, since the BAG relies on these scores to estimate exploitation probability, this re-scoring also directly impacts the posterior probability of node exploitation, refining the vulnerability assessment to better represent the likelihood of exploitation for each node. 

Looking to the future, this approach can incorporate additional data sources and other elements of the CVSS framework — such as Environmental Metrics — to further refine prioritization based on an organization’s specific technology stack, potential impact on operations, and existing security controls.


\section*{Acknowledgment}
This work was supported by Agenzia per la Cybersicurezza Nazionale under the programme for promotion of XL cycle PhD research in cybersecurity – C83C24000790001. The views expressed are those of the authors and do not represent the funding institution.


\begin{thebibliography}{1}

\bibitem{CVSS_original}
Mell, P., Scarfone, K., \& Romanosky, S. (2006). Common vulnerability scoring system. \textit{IEEE Security \& Privacy, 4(6)}, 85-89.

\bibitem{CVSS_docs}
Common Vulnerability Scoring System v3.1: Specification Document. FIRST. [Online]. Available: https://www.first.org/cvss/v3.1/specification-document

\bibitem{guarino1}
Guarino, S., Vitale, F., Flammini, F., Faramondi, L., Mazzocca, N., \& Setola, R. (2023). A two-level fusion framework for cyber-physical anomaly detection. \textit{IEEE Transactions on Industrial Cyber-Physical Systems, 2}, 1-13.

\bibitem{guarino3}
Guarino, S., Ansaldi, S., \& Setola, R. Multiple-Bayesian-Network-Based Risk Assessment Methodology for Industrial Control Systems. In \textit{Critical Infrastructure Protection XVIII: 18th IFIP WG 11.10 International Conference, ICCIP 2024, Arlington, VA, USA, March 18–19, 2024, Proceedings} (p. 113). Springer Nature.


\bibitem{laitila}
Laitila, P., \& Virtanen, K. (2016). Improving construction of conditional probability tables for ranked nodes in Bayesian networks. \textit{IEEE Transactions on Knowledge and Data Engineering, 28(7)}, 1691-1705.

\bibitem{Gonzalez}
Munoz-González, L., \& Lupu, E. C. (2016). Bayesian attack graphs for security risk assessment. In \textit{IST-153 Workshop on Cyber Resilience}.

\bibitem{williams}
Williams, T. J. (1994). The Purdue enterprise reference architecture. \textit{Computers in industry, 24(2-3)}, 141-158.

\bibitem{george}
George, P. G., \& Renjith, V. R. (2021). Evolution of safety and security risk assessment methodologies towards the use of bayesian networks in process industries. \textit{Process Safety and Environmental Protection, 149}, 758-775.

\bibitem{huang}
Huang, K., Zhou, C., Tian, Y. C., Tu, W., \& Peng, Y. (2017, November). Application of Bayesian network to data-driven cyber-security risk assessment in SCADA networks. In \textit{2017 27th International Telecommunication Networks and Applications Conference (ITNAC)} (pp. 1-6). IEEE.


\bibitem{guarino2}
Guarino, S., Faramondi, L., Oliva, G., Del Prete, E., \& Setola, R. (2024, June). Holistic Risk Assessment in Industrial Control Systems: Combining Multiple Bayesian Networks with Multi-Criteria Decision Making. In \textit{2024 32nd Mediterranean Conference on Control and Automation (MED)} (pp. 37-42). IEEE.

\bibitem{Poolsappasit}
Poolsappasit, N., Dewri, R., \& Ray, I. (2011). Dynamic security risk management using bayesian attack graphs. \textit{IEEE Transactions on Dependable and Secure Computing, 9(1)}, 61-74.

\bibitem{singh}
Singh, U. K., \& Joshi, C. (2016, October). Quantitative security risk evaluation using CVSS metrics by estimation of frequency and maturity of exploit. In \textit{Proceedings of the World Congress on Engineering and Computer Science} (Vol. 1, pp. 19-21).

\bibitem{meng}
Meng, H., An, X., \& Xing, J. (2022). A data-driven Bayesian network model integrating physical knowledge for prioritization of risk influencing factors. \textit{Process Safety and Environmental Protection, 160}, 434-449.

\bibitem{sato}
Sato, R., Kawaguchi, H., \& Nakatani, Y. (2022, December). A Stochastic Model for Calculating Well-Founded Probabilities of Vulnerability Exploitation. In \textit{2022 IEEE 22nd International Conference on Software Quality, Reliability, and Security Companion (QRS-C)} (pp. 34-43). IEEE.


\end{thebibliography}
\end{document}